\documentclass[twocolumn,nofootinbib,showpacs]{revtex4}

\usepackage[utf8]{inputenc}
\usepackage[T1]{fontenc}
\usepackage[fleqn]{amsmath}
\usepackage{graphicx}
\usepackage{float}
\usepackage[font={small,it}]{caption}
\restylefloat{figure}
\include{amssymbol}
\input{epsf}

\def\to{\rightarrow}
\def\d{\partial}
\def\dota{\dot{a}}
\def\ddota{\ddot{a}}

\def\al{\alpha} \def\be{\beta}  
   
\def\th{\theta}   \def\ka{\kappa}
\def\la{\lambda}  \def\rh{\rho} 
   
\def\ph{\phi}   
  \def\De{\Delta} 
\def\La{\Lambda}   
 \def\Om{\Omega} \def\mn{{\mu\nu}} \def\cl{{\cal L}}
\def\fr#1#2{{{#1} \over {#2}}}

\def\expect#1{\langle{#1}\rangle}

 \def\frac#1#2{{\textstyle{{#1}\over
{#2}}}}
\def\lsim{\mathrel{\rlap{\lower4pt\hbox{\hskip1pt$\sim$}}
\raise1pt\hbox{$<$}}}
\def\gsim{\mathrel{\rlap{\lower4pt\hbox{\hskip1pt$\sim$}}
\raise1pt\hbox{$>$}}} \def\sqr#1#2{{\vcenter{\vbox{\hrule height.#2pt
\hbox{\vrule width.#2pt height#1pt \kern#1pt \vrule width.#2pt} \hrule
height.#2pt}}}}
\def\square{\mathchoice\sqr66\sqr66\sqr{2.1}3\sqr{1.5}3}
\def\beq{\begin{equation}} \def\eeq{\end{equation}}
\def\beqa{\begin{eqnarray}} \def\eeqa{\end{eqnarray}}
\def\eq#1{Eq. (\ref{#1})}

\begin{document}

\title{Cosmological implications of bumblebee vector models}

\author{Diogo Capelo}
\email{diogo.capelo@ist.utl.pt}
\affiliation{Departamento de F\'isica, Instituto Superior T\'ecnico, Universidade de Lisboa,\\Av. Rovisco Pais 1, 1049-001 Lisboa, Portugal}

\author{Jorge P\'aramos}
\email{jorge.paramos@fc.up.pt}
\affiliation{Centro de F\'isica do Porto and Departamento de F\'{\i}sica e Astronomia, Faculdade de Ci\^encias da Universidade do Porto,\\Rua do Campo Alegre 687,
4169-007 Porto, Portugal}

\date{\today}

\begin{abstract}
The bumblebee model of spontaneous Lorentz symmetry breaking is explored in a cosmological context, considering a single nonzero time component for the vector field. The relevant dynamic equations for the evolution of the Universe are derived and their properties and physical significance studied. We conclude that a late-time de Sitter expansion of the Universe can be replicated, and attempt to constrain the parameter of the potential driving the spontaneous symmetry breaking.
\end{abstract}


\maketitle 

\section{Introduction}

Several outstanding questions in cosmology, such as the current accelerated expansion of the Universe \cite{Riess:1998cb} or the presence and origin of dark matter \cite{Trimble:1987ee}, have motived many efforts to expand upon the conventional formalism of general relativity (GR), as the latter cannot provide a direct explanation of such phenomena. A great number of attempts have been made to phenomenologically study putative modifications of GR, ranging from the inclusion of nontrivial curvature terms in the Einstein-Hilbert action \cite{DeFelice:2010aj,Bertolami:2013xda} to the addition of suitable scalar \cite{Zlatev:1998tr} and vector fields \cite{Tartaglia:2007mh,ArmendarizPicon:2009ai,aether,kostelecky1}, among others \cite{Clifton:2011jh}. 

Amongst the latter, the bumblebee model was initially put forwarded in 1989 \cite{kostelecky0} as a toy model for spontaneous Lorentz symmetry breaking using a vector field; being an extension of GR, it takes the usual Einstein-Hilbert action and expands it with a vector field and a potential. It provided a simple and more tractable scenario than the more general Standard Model extension, a framework for studying Lorentz-breaking terms added to the Standard model of interactions and general relativity \cite{kostelecky2}; several aspects of the impact of the SME on the gravitational sector have been addressed {\it e.g.} in Refs. \cite{SMEg1,SMEg2,SMEg3,SMEg4,SMEg5,SMEg6}.

The bumblebee model can be considered as a particular case of vector-tensor models \cite{Tartaglia:2007mh,ArmendarizPicon:2009ai}, which also include the more well-known Einstein-aether models \cite{aether}; in the latter, the potential of the vector field resorts to a Lagrange multiplier that constrains its norm, and exhibits couplings between the vector field and the Ricci tensor and scalar. Distinctively, the bumblebee model assumes only that the potential has a nonvanishing vacuum expectation value (VEV), but restricts the coupling to just the Ricci tensor, as depicted in the action functional (\ref{eqAction}).

The bumblebee model incorporates a mechanism of spontaneous breaking of the Lorentz symmetry, which is compatible with general Riemann-Cartan geometries \cite{kostelecky1,Bailey:2006fd,kostelecky2,Seifert:2009gi,Bertolami:2005bh,Paramos:2014mda}. This mechanism is inspired by the Higgs mechanism of the Standard model of fundamental particles and interactions (itself an allusion to the Landau theory of phase transitions that first originated in Condensed Matter Physics), and serves to both safeguard Lorentz symmetry as a symmetry of the action and regulate the way in which that symmetry is broken.

The coupling of curvature to the vector field and the presence of a potential leads to a higher complexity of the equation of motion that cannot be extrapolated from approaches that eschew the use of a potential \cite{JimenezMaroto,Zuntz} or uncouple the vector field from the curvature \cite{KoivistoMota}.

The aim of this work is to further assess the impact of the bumblebee model and explore the effects of a spontaneous LSB in a cosmological context --- namely its potential as a candidate for dark energy. 

This work is organized as follows: first, the model and its governing equations are described; the impact of the latter on cosmology is then presented, prompting the study of the ensuing dynamical system; finally, the obtained results are discussed and conclusions are drawn.

\section{The bumblebee model for Spontaneous Lorentz Symmetry Breaking}

As mentioned in the previous section, the bumblebee model extends the standard formalism of general relativity by allowing a LSB; this is dynamically driven by a suitable potential exhibiting a nonvanishing VEV, so that the bumblebee vector field $B_\mu$ acquires a specific four-dimensional orientation.

The action functional should capture the most relevant features of the bumblebee model, namely the coupling between the bumblebee field and geometry and the presence of a general potential. As discussed in detail in Ref. \cite{ArmendarizPicon:2009ai}, the simultaneous inclusion of kinetic terms for the vector field of the forms $ F_\mn F^\mn$ (with $F_\mn $ the field-strength) and $(\nabla_\mu B^\mu)^2$ leads to the appearance of ghost degrees of freedom, which vanish if only one of the latter is considered.

Furthermore, an extensive discussion of the symmetries of a model with couplings of the form $B^\mu B^\nu R_\mn $ and $ B_\mu B^\mu R $ was presented in Ref. \cite{SMEg2}: in particular, if the potential is quadratic and includes a Lagrange multiplier $\lambda$, $V(\la,B_\mu) = \la (B_\mu B^\mu \pm b^2) $, so that variation with respect to the former fixes $B_\mu B^\mu = \pm b^2$, then the coupling term $ B_\mu B^\mu R $ can be gauged away by rescaling $G$.

This said, the fixed points obtained in this study indicate that the bumblebee field does not rest at its nonvanishing VEV, for a general potential, so that the inclusion of both types of couplings should have physical consequences. However, for simplicity we drop the coupling to the Ricci scalar and thus consider the model posited in Ref. \cite{kostelecky1}, thus obtaining a more computationally tractable problem which still retains the main features mentioned above; a discussion of the effect of an additional coupling with the Ricci scalar can be found in the Appendix, where it is shown that --- although the strength of the additional coupling could lead to new fixed points --- the overall dynamical structure of the field equations is not substantially modified.

Given the above, we consider the action functional
\beq
\begin{split}
S = & \int \sqrt{-g} \left[ \fr{1}{2\ka} \left( R + \xi B^\mu B^\nu R_{\mu\nu} \right) \right.- \\
    & - \left. \fr{1}{4}B^{\mu\nu} B_{\mu\nu} - V \left( B^\mu B_\mu \pm b^2 \right) + \mathcal{L}_M \right] d^4 x~~,
\end{split}
\label{eqAction}
\eeq
\noindent where $\ka = 8\pi G$, $\xi$ is a coupling constant (with dimensions $[\xi ] =M^{-2}$), $B_\mu$ is the bumblebee field (with $[B_\mu] = M$), $B_\mn\equiv \d_\mu B_\nu - \d_\nu B_\mu$ is the field-strength tensor, $b^2 \equiv b_\mu b^\mu = \expect{B_\mu B^\mu}_0 \neq 0$ is the expectation value for the contracted bumblebee vector, $V$ is a potential exhibiting a minimum at $B_\mu B^\nu \pm b^2 = 0$, and $\mathcal{L}_M$ is the Lagrangian density for the matter fields.

By varying Eq. (\ref{eqAction}) with respect to the metric, the modified Einstein equations are obtained,
\beqa \label{eqEinstein}
\nonumber G_\mn &=& \ka \bigg[ 2 V' B_\mu B_\nu- B_{\mu\al}B^\al_{\nu} - \left( V + {1 \over 4} B_{\al\be}B^{\al\be}\right) g_\mn \bigg] \\ \nonumber && + \xi  \bigg[ {1 \over 2} B^\al B^\be R_{\al\be} g_\mn - B_\mu B^\al R_{\al\nu} - B_\nu B^\al R_{\al\mu}  \\ \nonumber && +{1 \over 2}  \nabla_\al \nabla_\mu (B^\al B_\nu) + {1 \over 2}  \nabla_\al \nabla_\nu (B^\al B_\mu)  \\ && - {1 \over 2}  \nabla_\al \nabla_\be( B^\al B^\be) g_\mn - {1 \over 2} \square ( B_\mu B_\nu ) \bigg] + T_\mn ~~,
\eeqa
\noindent where $V'$ denotes the derivative of the potential $V$ with respect to its argument and $T_\mn$ is the energy-momentum tensor of matter. 

Variation of Eq. (\ref{eqAction}) with respect to the bumblebee field yields its equation of motion,
\beq
\nabla_\mu B^\mn = 2   \left( V' B^\nu - {\xi \over 2\ka}   B_\mu R^\mn\right)~~.
\label{eqbumblebee}
\eeq
If the lhs of the equation vanishes, the above results in a simple algebraic relation between the bumblebee, its potential and the geometry of spacetime.

\section{Cosmology}
\label{cosmology}

In most cosmological studies, the Friedmann-Robertson-Walker metric (FRW) metric is assumed, reflecting the cosmological principle which posits an homogeneous and isotropic Universe. However, if Lorentz symmetry is spontaneously broken, it is possible that the bumblebee field acquires a nonvanishing spatial orientation, which would dynamically break the aforementioned isotropy and require a more evolved geometry. This possibility (elaborated in the final section) shall not be pursued in this study: instead, one considers the ansatz for the bumblebee field,
\beq \label{eqBform} B_\mu = \left(B(t),  \vec{0}\right)~~, \eeq
\noindent thus upholding the validity of the cosmological principle, {\it i.e.} maintaining the assumption of large-scale homogeneity and isotropy of the Universe. One thus adopts the flat FRW metric, as given by the line element,
\beq
ds^2 = - dt^2 + a(t)^2 \left[ dr^2 + r^2   d\th^2 + r^2 \sin^2(\th)   d\ph^2 \right] \label{FRW}~~,
\eeq
\noindent where $a(t)$ is the scale factor.

From Eq. \ref{eqBform}, one sees that the field-strength tensor vanishes, $B_\mn = 0$, and the only nontrivial component of the bumblebee Eq. (\ref{eqbumblebee}) is
\beq \left( V' - {3 \xi \over 2 \ka } {\ddota \over a}\right) B = 0 ~~,
\label{eqVprime}
\eeq
\noindent which, for a nonvanishing bumblebee field, establishes a relation between the dynamics of the potential and the scale factor. This is probably the reason that motivated the authors of Ref. \cite{ArmendarizPicon:2009ai} to describe bumblebee models with only a Maxwell kinetic term as ``non-dynamical''; we maintain, however, that the nonminimal coupling of curvature in our case is enough to drive a meaningful process, as shall be studied below.

Considering the matter energy-momentum tensor for pressureless dust, 
\beq T_\mn = \rho u_\mu u_\nu ~~, \eeq
\noindent where $\rho $ is the energy density of matter, $u_\mu $ is the four-velocity (with $u_\mu= (1,\vec{0}) $ due to the normalization condition $u_\mu u^\mu = -1$), and using the above expression together with  Eq. (\ref{eqBform}), the $t-t$ component of the modified field Eqs. (\ref{eqEinstein}) becomes
\beq H^2 (1-\xi B^2) = {1 \over 3} \ka (\rho + V ) + \xi H B \dot B ~~,
\label{eqEinstTT}
\eeq
\noindent while the diagonal $i-i$ components read
\beqa \label{eqEinstRR} && \left( H^2  + 2 {\ddot{a} \over a} \right) ( 1 - \xi B^2) = \\ \nonumber && \ka V +  \xi \left(4HB\dot B + \dot B ^2 + B \ddot B \right) ~~, \eeqa
\noindent where $H \equiv \dot{a}/a$ is the Hubble parameter.

Using the Bianchi identities, one also obtains the following modified equation of conservation of energy,
\beq \dot{\rh} = - 3H\rho - 3{ \xi \over \ka} {\ddot{a}\over a} B ( HB + \dot{B}) + 3{ \xi \over \ka}{\dddot{a}\over a} B^2 ~~,
\label{eqConsE}
\eeq
\noindent showing that there is an energy exchange between matter and the bumblebee field.

Due to the complexity of the equations derived in this section, a simple, closed-form solution cannot be obtained by purely analytical means. Given this difficulty, it becomes more enlightening to study instead the full dynamical picture presented by the Friedmann and Raychaudhuri equations under the constraints imposed by the bumblebee field equation of motion.

\subsection{Absence of coupling between the bumblebee field and the Ricci tensor}
\label{SSECdesitter}

At this point, it is useful no notice the relevance of the coupling strength $\xi$: if this is ``switched off'' by setting $\xi = 0$, the modified Einstein Eqs. (\ref{eqEinstein}) become
\beqa G_\mn &=& T_\mn + \\ \nonumber && \ka \bigg[ 2 V' B_\mu B_\nu- B_{\mu\al}B^\al_{\nu} - \left( V + {1 \over 4} B_{\al\be}B^{\al\be}\right) g_\mn \bigg] ~~, \label{eqEinstein_no} \eeqa
\noindent while the bumblebee equation of motion (\ref{eqbumblebee}) reads 
\beq
\nabla_\mu B^\mn = 2 V' B^\nu ~~.
\label{eqbumblebee_no}
\eeq
Although simplified, the above equations in principle still allows for dynamical behavior. However, the assumption \eq{eqBform} leads to a vanishing field force $B^\mn = 0$, so that the bumblebee Eq. (\ref{eqVprime}) collapses to
\beq V' B = 0 ~~, \eeq
\noindent forcing the bumblebee field to either vanish or rest at one of the extrema of its potential, and keeping it from evolving with time.
 
Thus, setting $\xi = 0 $ and $\dot{B} = \ddot{B} = 0 $ in Eqs. (\ref{eqEinstTT}-\ref{eqConsE}) yields the simplified set,
\beqa \label{Friedmann_no} H^2 &=& {1 \over 3} \ka (\rho + V_0 )  ~~, \\  \label{Raychaudhuri_no}  H^2 + 2 {\ddota \over a} &=& \ka V_0 ~~, \\ \label{cons_no} \dot{\rh} &=& - 3H\rho ~~.\eeqa
Thus, the bumblebee field contributes solely through the value of its constant potential $V(B^2 ) \equiv V_0$: checking the action functional (\ref{eqAction}), one sees that it collapses to the Einstein-Hilbert form with a Cosmological Constant $\La = \ka V_0$. Indeed, \eq{Raychaudhuri_no} can be directly integrated, yielding a de Sitter solution
\beq
a(t) = a_0   e^{H_0 (t-t_0)}~~.
\label{eqAdeSitter}
\eeq
\noindent with $H_0^2 \equiv \ka V_0 /3 = \La/3$; \eq{Friedmann_no} then requires that $\rho = 0$, as expected.

As the next section will show, the existence of a phase of exponential acceleration of the Universe is not restrained to the uncoupled scenario $\xi = 0$, but for a nonvanishing coupling as well.

\subsection{Generality of a de Sitter solution}
\label{SSECdesitter}

Following the preceding discussion, we consider a solution of the form \eq{eqAdeSitter} (with unknown $H_0$), so that the bumblebee field equation (\ref{eqVprime}) becomes
\beq
V'  B^2 = {3 \over 2} {\xi \over \ka} {H_0}^2  B ^2~~.
\label{eqVpEqConst}
\eeq
\noindent For a nonvanishing bumblebee field, any potential $V$ which varies nonlinearly yields a quantity $V'$ that depend on $B$, and the above implicitly establishes that the bumblebee field must be constant through $V'(B^2\pm b^2) = (3 \xi / 2\ka) H_0^2 $ (if $B=0$ this relation does not hold, but the bumblebee remains constant nevertheless). Using $B=B_0 = {\it const}$ on Eqs. (\ref{eqEinstTT}) and (\ref{eqConsE}) yields
\beq \label{eqHVBsitter} H_0^2 = {\ka V_0 \over 3(1-\xi B_0^2)} ~~~~,~~~~\dot{\rho} = -3 H_0 \rho = 0 ~~. \eeq
\noindent As in the previous section, this corresponds to a Universe without matter and dominated by the bumblebee field alone, which acts as a form of dark energy.

Given a particular potential $V$, Eq. (\ref{eqVpEqConst}) and the above fully determine both $H_0$ and the nonvanishing value of $B_0$. Anticipating the next study, one considers a power-law potential of the form
\beq
V (B^2 \pm b^2)= M^{4-2n}( B ^2 \pm b^2)^n\rightarrow V' = {n V \over  B ^2 \pm b^2}~~,
\label{eqVpl}
\eeq
\noindent where $M$ has dimensions of mass. Using Eqs. (\ref{eqHVBsitter}) and (\ref{eqVpEqConst}) for $B \neq 0$, one obtains
\beqa \label{dSnonvanFP}
B_0^2 &=& { 2n \mp \xi b^2 \over \xi (1+2n)} \rightarrow \\ \nonumber V_0&=& \left( {2n\over 1+2n}{1 \pm \xi b^2 \over \xi M^2} \right)^n M^4 \rightarrow \\ \nonumber H_0^2 &=& {2 n \over 3} {\ka \over \xi} \left( { 2n \over 1+2n}{ 1\pm \xi b^2 \over \xi M^2 } \right)^{n-1} M^2~~.  
\eeqa
\noindent If the bumblebee field vanishes, then one trivially obtains $H_0^2 = \ka V_0/3$.

Notice that in the above one does not simply find that $ B^2 = \mp b^2$ and $V(B^2 \pm b^2 ) = V(0) = 0$, as could be naively expected if one assumed that the cosmological dynamics should evolve the bumblebee field until it rests at its nonvanishing VEV $<B^2 > = b^2$: indeed, following the discussion of the previous section, the dynamical effect of the coupling term $\xi B^\mu B^\nu R_\mn $ found in the action functional (\ref{eqAction}) on the field equations can be interpreted as a friction term that arrests the evolution of the bumblebee field, with the dissipated energy acting to counteract the gravitational attraction of matter and drive an accelerated expansion of the Universe.

One can now look back into the on-shell form of the action (\ref{eqAction}), which --- with the prescription (\ref{eqBform}), $B(t) =B_0$ and the de Sitter solution (\ref{eqHVBsitter}) --- is reduced to
\beqa
S &=& \int  {1 \over 2\ka} \left[ R  - 6 \left(1 - {\xi B_0^2\over 2} \right) {H_0}^2  \right] \sqrt{-g} d^4 x~~,\eeqa
\noindent which, as noted in the preceding section, corresponds to the Einstein-Hilbert action with an added constant term which acts as a cosmological constant $\La$: if the bumblebee field vanishes, the latter is given by its usual definition $\La = 3H_0^2$, by comparison with the usual Lagrangian density $\cl = (R - 2\La)/(2\ka)$; if one considers $B_0\neq 0$, then it also includes a contribution from the coupling between the bumblebee field and the Ricci tensor, $\La^* \equiv 3H_0^2(1-\xi B_0^2/2)$ (see Ref. \cite{Ribeiro:2014sla} for a study of the contribution to a phase of accelerated expansion of the Universe arising from a nonminimal coupling between matter and the Ricci scalar).

\section{Analysis of the Dynamical System}

In order to explore the possible solutions to Eqs. (\ref{eqEinstTT}-\ref{eqConsE}) and confirm the results obtained in the previous section, one begins by defining the following dimensionless variables (see Refs. \cite{Tsujikawa:2010sc} for a similar treatment in the context of quintessence),
\beq
x_1 = {\ka V \over 3 H^2} ~~,~~x_2 = \xi  B ^2~~,~~x_3 = {\xi  B \dot B \over H}~~,
\eeq
\noindent together with the usual relative matter density and deceleration parameter
\beq
\Om_M = {\ka \rh \over 3 H^2} ~~~~,~~~~q = -{\ddota a \over \dota^2}~~.
\eeq
Using Eqs. (\ref{eqVprime}), (\ref{eqEinstTT}) and (\ref{eqConsE}), one obtains 
\beqa
x'_1 &=& 2 \left( 1 + \al x_3 + q \right) x_1 ~~, \label{DINsystS} \\  \nonumber
x'_2 &=& 2x_3 ~~, \\ \nonumber
x'_3 &=& ( 1 -2q ) (1 - x_2 ) - 3 x_1 + x_3( q-3) ~~,
\eeqa
\noindent where the prime denotes differentiation with respect to the number of {\it e-folds} $N \equiv \log a$ and one defines
\beq
\al(x_2) \equiv {V'(x_2) \over V(x_2)} ~~~~,~~~~ \be \equiv {\dddot{a} \over a H^3}~~.
\eeq
We notice that the dimensionless variable $x_2$ is directly proportional to $B^2$, so that no expansion is performed around the nonvanishing VEV $b^2 \neq 0$ arising after the LSB; conversely, this definition appears to indicate that the vector field is expanded around its symmetric phase $B^2 = 0 $. However, since the set of fixed points does not change with a shift in the definition of $x_2$, there is no distinguishable physics in adopting either form (this was checked directly by instead using the definition $x_2 = \xi (B^2 - b^2)$).

Furthermore, the calculations below reveal that no fixed point arises with $B^2 = b^2$ ({\it i.e.} $x_2 = \xi b^2$): the bumblebee field never rolls to its nonvanishing VEV, due to the effect of the coupling with the Hubble parameter (and its derivative) in the equations of motion.

The modified Friedmann equation (\ref{eqEinstTT}) reduces to
\beq
1 = x_1 + x_2 + x_3 + \Om_M~~,
\label{DINfried}
\eeq
\noindent while the bumblebee equation of motion becomes
\beq
(2\al x_1 + q)  x_2 = 0~~.
\label{DINbumb}
\eeq
The two equations above are algebraic constraints that must be obeyed by any solution to the dynamical system (\ref{DINsystS}).

The form of Eq. (\ref{DINbumb}) indicates two possible branches, corresponding to whether $x_2$ is vanishing or not. If the latter is true, this equation collapses to $q = -2\al x_1 $, which can then be used to further simplify the dynamical system (\ref{DINsystS}). The opposite case $x_2=0$ is more troublesome, as it implies that the dynamical system is no longer autonomous [in the sense that $x_i' = f(x_j)$ alone]: one would have to promote $q$ and $\Om_m$ to dynamical variables and consider the two additional differential equations
\beqa \label{DINsystA} \Om'_M &=& (2q-1)\Om_M  + \be x_2 - 2\al x_1 ( x_2 + x_3 ) ~~, \\ \nonumber 
q' & = & - \be + q + 2 q^2 ~~.\eeqa
\noindent Furthermore, it can be shown that the determination of the nature of any fixed points would be problematic, as the linearization of the Jacobian matrix of the system around $x_2=0$ diverges.

This caveat, however, may be circumvented if one notices that $x_2= 0$ is not dynamically relevant if $x'_2 \neq 0$, as the variable just rolls out of its vanishing value and the relation $q = -2\al x_1 $ becomes immediately valid, allowing the aforementioned substitution into Eqs. (\ref{DINsystS}).

Furthermore, since one is not interested in modeling the spontaneous LSB, it is natural to assume that the bumblebee field has had time to roll from its previous vanishing VEV (before the LSB) towards the broken phase $<B^2> = b^2$.

Thus, one is left only with the pathological case $x_2 = x_2' = 0$, which is physically hard to interpret, as it would imply that the spontaneous LSB has no dynamical effect on the Bumbleebee field, which forever rests at its pre-LSB vanishing VEV.

Fortunately, it can be solved analytically; since the physical implication of this is that the bumblebee field only imprints an effect on the dynamics through the constant value of the potential $V_0$, one expects to recover the usual picture found in GR for a Universe composed of matter and a cosmological constant, where the former becomes more and more diluted and the expansion grows exponentially at late times. 

Indeed, taking Eqs. (\ref{DINsystS}) and (\ref{DINsystA}) and replacing $x_2 = x'_2 = 0$ immediately yields $x_3=0$, so that
\beqa
x'_1 &=& 2 \left( 1 + q \right) x_1 ~~, \\  \nonumber
3 x_1  &=& 1 -2q ~~, \\ \nonumber
\Om'_M &=& (2q-1)\Om_M  ~~.
\eeqa
Notice that the last equation is equivalent to the first, as can be seen from the Friedmann equation ({\ref{DINfried}), which reads $\Om_m = 1 - x_1 \rightarrow \Om_m' = -x_1'$. One obtains the single differential equation
\beq x'_1 = 3 \left( 1-x_1 \right) x_1 ~~, \eeq
\noindent with solution
\beqa x_1(N) &=& {1 \over 1 + C e^{-3N}} \rightarrow \\ \nonumber \Om_m(N) &=& 1-x_1= { C \over  C + e^{3N}}  ~~, \\ \nonumber q(N) &=& {1-3x_1\over 2} = -1 + {3 \over 2} { C \over C +e^{3N}} ~~, \eeqa
\noindent where $C$ is an integration constant. As expected, one finds that $\Om_m $ becomes vanishingly small as the Universe expands and approaches a de Sitter phase $q=-1$. One concludes that, even if the bumblebee field becomes locked in the symmetric phase $<B^2> = 0$ so that $x_2= x'_2=0$, this does not give rise to any unphysical dynamics, and can proceed towards the study of the more relevant scenario $x_2 \neq 0$.

As discussed previously, a nonvanishing bumblebee field implies that $q = -2\al x_1$; replacing this into the dynamical system (\ref{DINsystS}) yields the closed set
\beqa \label{DINsystFx2}
x'_1 &=& 2 \left( 1 + \al x_3 -2\al x_1 \right) x_1 ~~,\\ \nonumber
x'_2 &=& 2x_3 ~~, \\ \nonumber
x'_3 &=& ( 1 +4\al x_1 ) (1 - x_2 ) - 3 x_1 - x_3(3 + 2\al x_1) ~~,
\eeqa
\noindent and the relation 
\beq x_1 + x_2 + x_3 + \Om_M= 1~~.\eeq
Following the results obtained in the previous section, one adopts a power-law potential of the form (\ref{eqVpl}), here rewritten as
\beq V(x_2) = M_*^4 \left( x_2 - \xi b^2\right)^n~~,
\label{DINpot}
\eeq
\noindent where the sign is fixed and $\xi b^2$ is allowed to have negative values. It is worthy of note that, for this potential, the variable $\al$ becomes:
\beq
\al = {n \over x_2 - \xi b^2}~~.
\label{eqalpha}
\eeq
The presence of this multiplicative term on two of the three dynamical equations (\ref{DINsystFx2}) renders its behavior very important in understanding the evolution of the overall system. Since Eq. (\ref{DINfried}) implies that, for the matter-dominated Universe $\Om_m = 1 \to x_1 + x_2 + x_3 = 0$, if $x_2 > \xi b^2 > 0$, $x_1$ will be positive by definition, as seen in Eq. (\ref{DINpot}). In that case, $x_3$ will have to be negative, causing $x_2$ to decrease until it crosses the value of $\xi b^2$ and $\al$ diverges (and with it, the trajectories in phase space).

Therefore, one must enforce $x_{2i} < \xi b^2$, where the former is the initial value of the dimensionless variable $x_2$: notice that, since one is studying the cosmological dynamics after the spontaneous LSB has occurred and the potential $V$ has settled into the form (\ref{DINpot}), this initial value is endowed with physical meaning, as it is related to the value acquired by the bumblebee field after the LSB (which is not modeled directly).

The inequality  $x_{2i} < \xi b^2$ will imply that $\al$ is always negative, in order to avoid the divergence of the system: as the value of $x_2$ gets closer to $\xi b^2$, $\al$ becomes larger and (as will be seen numerically), the evolution in phase-space becomes more abrupt; thus, it is expected that some distance must be allowed between the initial values of $x_2$ and $\xi b^2$ for sufficiently smooth trajectories in phase space. Additionally, since the value of $\xi b^2$ is not expected to be very large (since $\xi$ is the constant responsible for the Lorentz-violating curvature coupling and no evidence of such effects have yet been observed), the imposed restraint on the initial values $x_{2} < \xi b^2$ becomes very demanding.

Under this potential, the system exhibits the fixed points shown on
\ref{tablef2}: in particular, fixed points ${\bf x}_B$ and ${\bf x}_C$ correspond to the de Sitter solutions found in section \ref{SSECdesitter} --- the former corresponds to either a vanishing coupling $\xi$ or a bumblebee field frozen at its unbroken VEV $B^2 = 0$, and is thus an unnatural candidate for a de Sitter attractor; conversely, ${\bf x}_C$ displays an additional contribution due to the coupling between the bumblebee field and the Ricci tensor.

\begin{table}[!h]
\begin{tabular}{rp{0pt}lp{0pt}}
 && $(x_1,x_2,x_3, \Om_m, q)$ \\
\hline
${\bf x}_A$ && $\left(0,0,0,1,{1 \over 2} \right)$ \\
${\bf x}_B$ && $\left(1,0,0,0,-1 \right)$ \\
${\bf x}_C$ && $\left({1+\xi b^2 \over 1+2n},{2n-\xi b^2 \over 1+2n},0,0,-1 \right)$\\
${\bf x}_D$ && $\left(0,1,0,0,0 \right)$
\end{tabular}
\caption{\rm Fixed points of the model, Eq. (\ref{DINsystFx2}).}
\label{tablef2}
\end{table}

The fixed point ${\bf x}_A$ corresponds to a matter dominated Universe before the bumblebee-induced dark energy becomes relevant. This transition requires that this fixed point is repulsive, with ${\bf x}_B$ and/or ${\bf x}_C$ attractive.

The fourth fixed point, ${\bf x}_D$, is unphysical, as it corresponds to a Universe without matter and only a constant bumblebee field, which either expands or contracts linearly or is static (so that $q=0$); although it is not mandatory for this fixed point to be repulsive (as it suffices that our Universe undergoes a trajectory in phase space sufficiently far from it), it is a desirable feature.

The analysis of the stability of the system can be achieved from studying the eigenvalues of the Jacobian matrices pertaining to the linearised system in each fixed point, which are:
\beq
D_A = \begin{bmatrix} 2 & 0 & -3-{4n \over \xi b^2} & {4n \over \xi b^2} \\ 0 & 0 & -1 & 0 \\ 0 & 2 & -3 & 0 \\ 0 & 0 & 0 & -1 \end{bmatrix}~~,
\eeq
\beq
D_B = \begin{bmatrix} 2+{8n \over \xi b^2} & 0 & -3-{4n \over \xi b^2} & 0 \\ {4n \over \xi b^2} & 0 & -1+{4n\left(\xi b^2-1\right) \over \xi b^2} & 0 \\ -{2n \over \xi b^2} & 2 & -3+{2n \over \xi b^2} & {2n \over \xi b^2} \\ 0 & 0 & 0 & -1+{4n \over \xi b^2} \end{bmatrix}~~,
\eeq
\beq
D_C = \begin{bmatrix} 2 & 0 & -1 & 0 \\ {1 \over n} & 0 & -3-{1 \over n} & 0 \\ 1 & 2 & -4 & {-2n^2+3n-2n\xi b^2 \over n\xi b^2-n}  \\ 0 & 0 & 0 & -3 \end{bmatrix}~~,
\eeq
\beq
D_D = \begin{bmatrix} 2 & 0 & -3 & 0 \\ 0 & 0 & -1 & 0 \\ 0 & 2 & -3 & 0 \\ 0 & 0 & 0 & -1 \end{bmatrix}~~,
\eeq
The eigenvalues $\la_i$ for matrices $D_A$ and $D_D$ do not depend on $n$ or $\xi b^2$:
\beqa
\la_A &=& \left( -2, -1, -1, 2 \right)~~, \\ \la_D &=& \left( -2, 2, -1, -1 \right)~~,
\eeqa
\noindent and, since at least one of their components has a positive real part, the two associated fixed points are unstable. For the remaining two fixed points, the issue of stability will vary, depending on the values taken by $n$ or $\xi b^2$. The specific form of the dependence is quite complex, but it can be understood in a fairly simple way by defining the third degree polynomials (of variable $\la$):
\beqa
p_B(\la) & = & \la^3 + \xi b^2 \left(\xi b^2 - 10 n\right) \la^2 + \\ \nonumber &&
2 \left( \xi b^2 \right)^2 \left[ 4n + 4n^2 - 17n \xi b^2 - 2 \left( \xi b^2 \right)^2 \right] \la - \\ \nonumber &&
4 \left( \xi b^2 \right)^3 \left[8n^2 - 2n\xi b^2 - 16n^2 \xi b^2 + \left( \xi b^2 \right)^3 \right]~~,
\eeqa
\beqa
p_C(\la) & = & \la^3 + 12 n^3 \left( \xi b^2 - 1 \right)^3 + \\ \nonumber && 2 n \left( \xi b^2 - 1 \right) \left( 3\la + \xi b^2 - 1 \right) \la + \\ \nonumber && 3 n^2 \left( \xi b^2 - 1 \right)^2 \left( 5\la + 2\xi b^2 - 2 \right)~~.
\eeqa
Denoting the three roots of each polynomial $r_{B,1}$, $r_{B,2}$, $r_{B,3}$ and $r_{C,1}$, $r_{C,2}$, $r_{C,3}$ respectively, the eigenvalues for $D_B$ and $D_C$ become:
\beq \la_B = \left( {4n \over \xi b^2} - 1, {r_{B,1} \over \left(\xi b^2\right)^2}, {r_{B,2} \over \left(\xi b^2\right)^2}, {r_{B,3} \over \left(\xi b^2\right)^2} \right)~~,\eeq
\noindent and
\beq \la_C = \left( -3, { r_{C,1} \over n \left(\xi b^2 - 1\right) }, { r_{C,2} \over n \left(\xi b^2 - 1\right) }, { r_{C,3} \over n \left(\xi b^2 - 1\right) } \right)~~. \eeq
\noindent The constraints on $n$ and $\xi b^2$ imposed by the stability of the fixed points ${\bf x}_B$ and ${\bf x}_C$ are shown in Fig. \ref{FIG_pars}.

\begin{figure}[!h]
\includegraphics[scale=0.4]{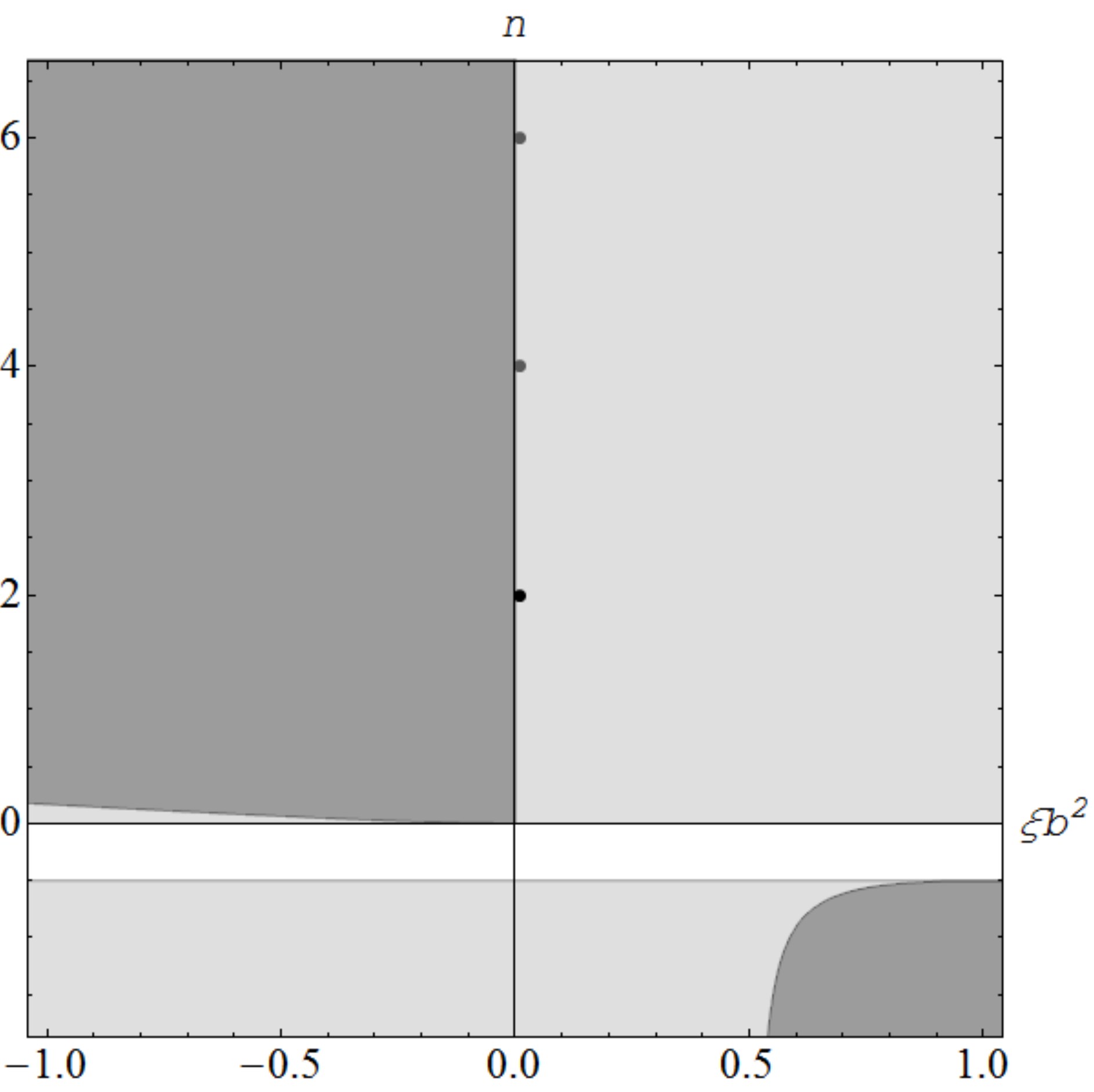}
\caption{Parameter constraints assuming ${\bf x}_B$ and ${\bf x}_C$ are unstable (white), only ${\bf x}_C$ is stable (light grey) and both ${\bf x}_B$ and ${\bf x}_C$ are stable (dark grey). The points used in the analysis of phase-space trajectories are highlighted.}
\label{FIG_pars}
\end{figure}

It can be seen immediately that only the region $-1/2<n<0$ warrants the instability of all of the system's fixed points, regardless of the value of $\xi b^2$. Additionally, if $n$ and $\xi b^2$ have the same sign, then only one stable fixed point is to be found (${\bf x}_C$); this fact being especially relevant since the simpler and more readily interpretable cases are those for which $n$ is positive (and small).

If one allows $\xi b^2$ to assume negative values while $n$ remains positive, then the two possible cases of only ${\bf x}_C$ being stable or both ${\bf x}_B$ and ${\bf x}_C$ being stable are separated by a line that can be roughly described by the equation $\xi b^2 \approx -4.76 n$, with the former lying below the line and latter, above. What this means is that to ensure the stability of  ${\bf x}_B$, $n$ must increase with $\xi b^2$. A similar behavior can be noted if, on the other hand, $n$ assumes negative values (lower than $-1/2$) and $\xi b^2$ is positive, although in this case the  equation describing the dividing line appears shifted (closer to $\xi b^2 \approx -4.76 n - 2.38$) and the area in which both fixed points are stable only asymptotically approaches the axis at $\xi b^2 = 0$. Again, this is  worthy of note since, as mentioned before, $\xi b^2$ is expected to be small, $\xi b^2 \ll 1$.

\section{Trajectory and Evolution Analysis}

Following the previous results, one now studies the trajectories of the variables in the phase-space, especially $x_2$, $\Om_M$ and $q$, {\it i.e.} the bumblebee field, matter density and deceleration parameter, respectively - which better provide a physical picture of the conditions of the Universe at the given time. One thus numerically integrates the system ({\ref{DINsystFx2}), with initial values corresponding to the departure from the matter dominated era, $\Om_M \approx 1$ and $q \approx 0.5$ (the evolution is not affected by choosing initial conditions sufficiently close to these values). A note should be made concerning the initial value of $x_2$, or ${x_2}_i$. The default would be to start the numerical simulation at $x_2=0$, but since the bumblebee field is in that position only before symmetry is broken, it is reasonable to expect a deviation from the origin, albeit small.

\subsection{Example of system convergence}
The case where $n=4$ is depicted in Figs. \ref{FIG2_45}-\ref{FIG2_24}, showing that the system collapses entirely to the single attractor, as determined by the constraints displayed in Fig. (\ref{FIG_pars}).

\begin{figure}[!h]
\includegraphics[scale=0.4]{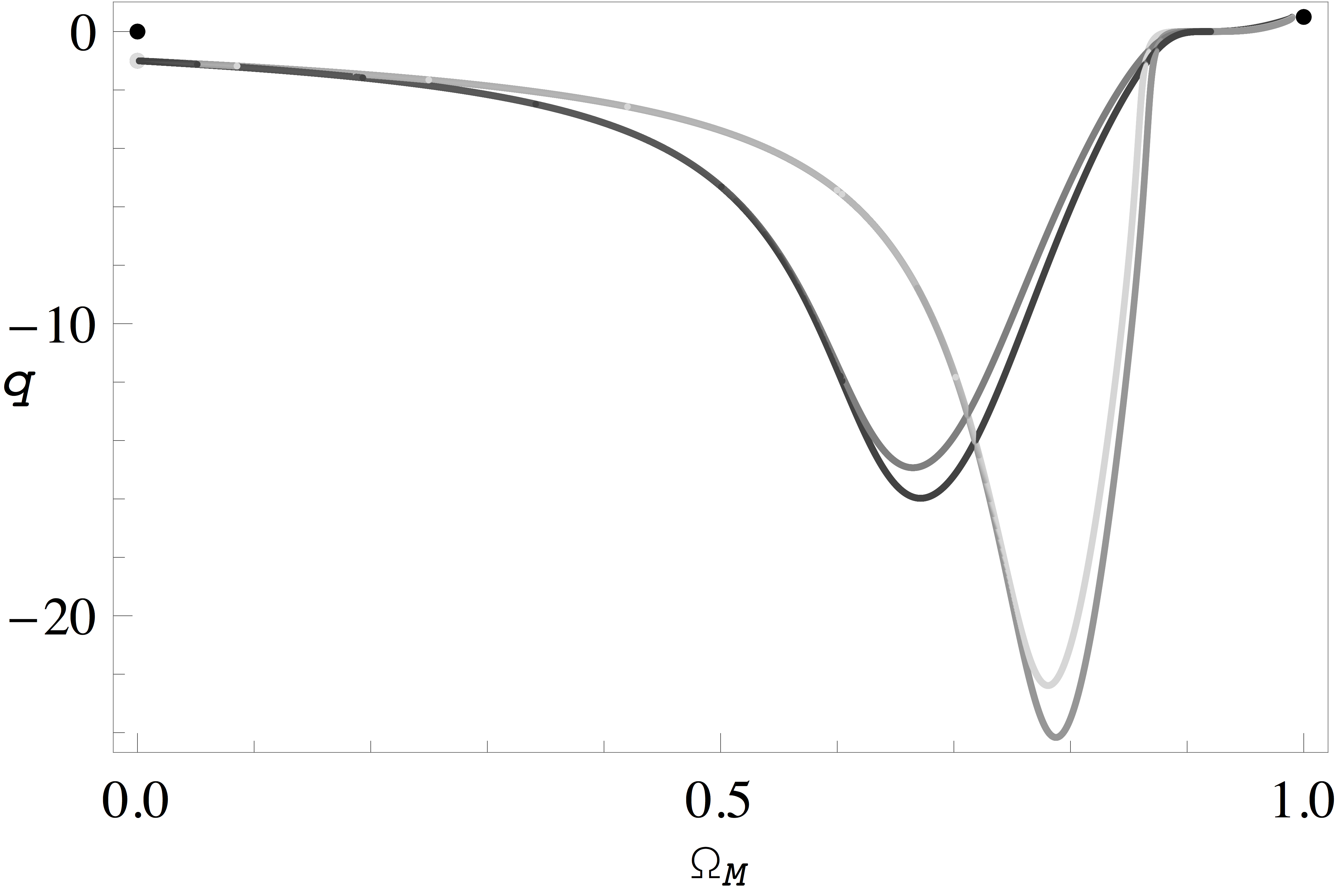}
\caption{Projection of the phase-space into the $\Om_M$-$q$ plane for $n=4$ (two darker plots) and $n=6$ (two lighter plots). Fixed points are marked as dots in dark grey ($n=4$) and light grey ($n=6$). For both cases $\xi b^2 = 10^{-2}$ and $x_{2i}=10^{-3}\;\text{(darker)},10^{-12}\;\text{(lighter)}$.}
\label{FIG2_45}
\end{figure}

\begin{figure}[!h]
\includegraphics[scale=0.4]{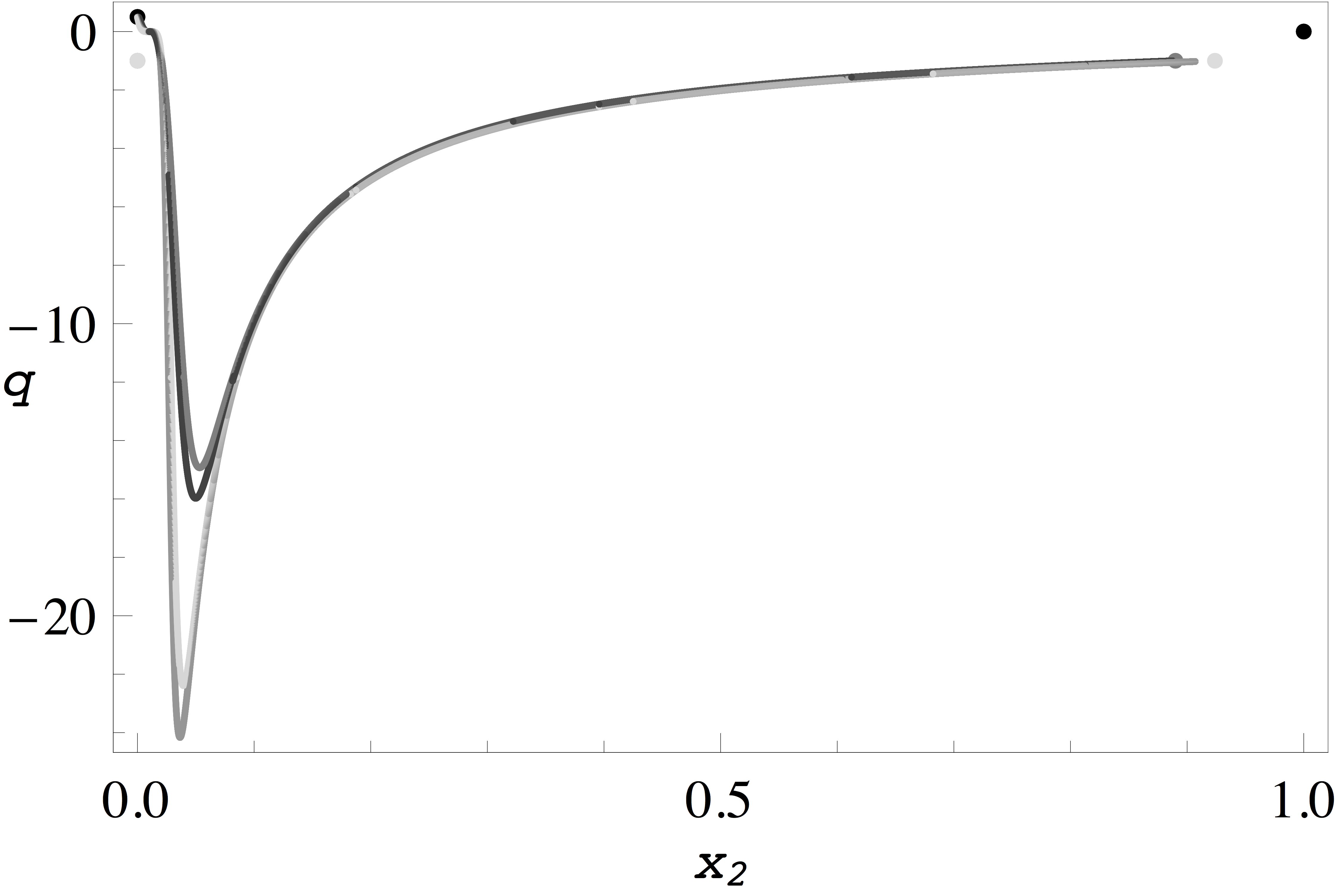}
\caption{Projection of the phase-space into the $x_2$-$q$ plane for $n=4$ (two darker plots) and $n=6$ (two lighter plots). Fixed points are marked as dots in dark grey ($n=4$) and light grey ($n=6$). For both cases $\xi b^2 = 10^{-2}$ and $x_{2i}=10^{-3}\;\text{(darker)},10^{-12}\;\text{(lighter)}$.}
\label{FIG2_25}
\end{figure}

These trajectories describe a Universe that transitions from an initial stage of matter domination ($\Om_M \approx 1, q \approx 0.5$) to a dark energy dominated era  ($\Om_M \approx 0, q \approx -1$). It is interesting to note that the behavior indicates that --- instead of evolving smoothly and monotonically towards a de Sitter phase --- the system overshoots that mark and crosses below $ q < -1$, before asymptotically approaching $q=-1$. Since the current value of $\Om_M \approx 0.3$ is well before that overshoot and cosmographic surveys indicate that $q$ has been steadily declining \cite{Gong:2006gs}, it is reasonable to assume that the lowest value of $q$ lies still in the future.

Furthermore, because of the discussion surrounding Eq. (\ref{eqalpha}), it can be seen that the restrictions imposed on the initial values of $x_2$ drive this overshoot deeply into negative values, $ q \ll -1$: mathematically, this issue could be settled by allowing $x_2 < 0$, which allows for much smoother trajectories (if still overshooting $q=-1$), but results on an added strain on the restriction Eq. (\ref{DINfried}), causing $\Om_M$ to dip into unphysical (negative) values.

\begin{figure}[!h]
\includegraphics[scale=0.4]{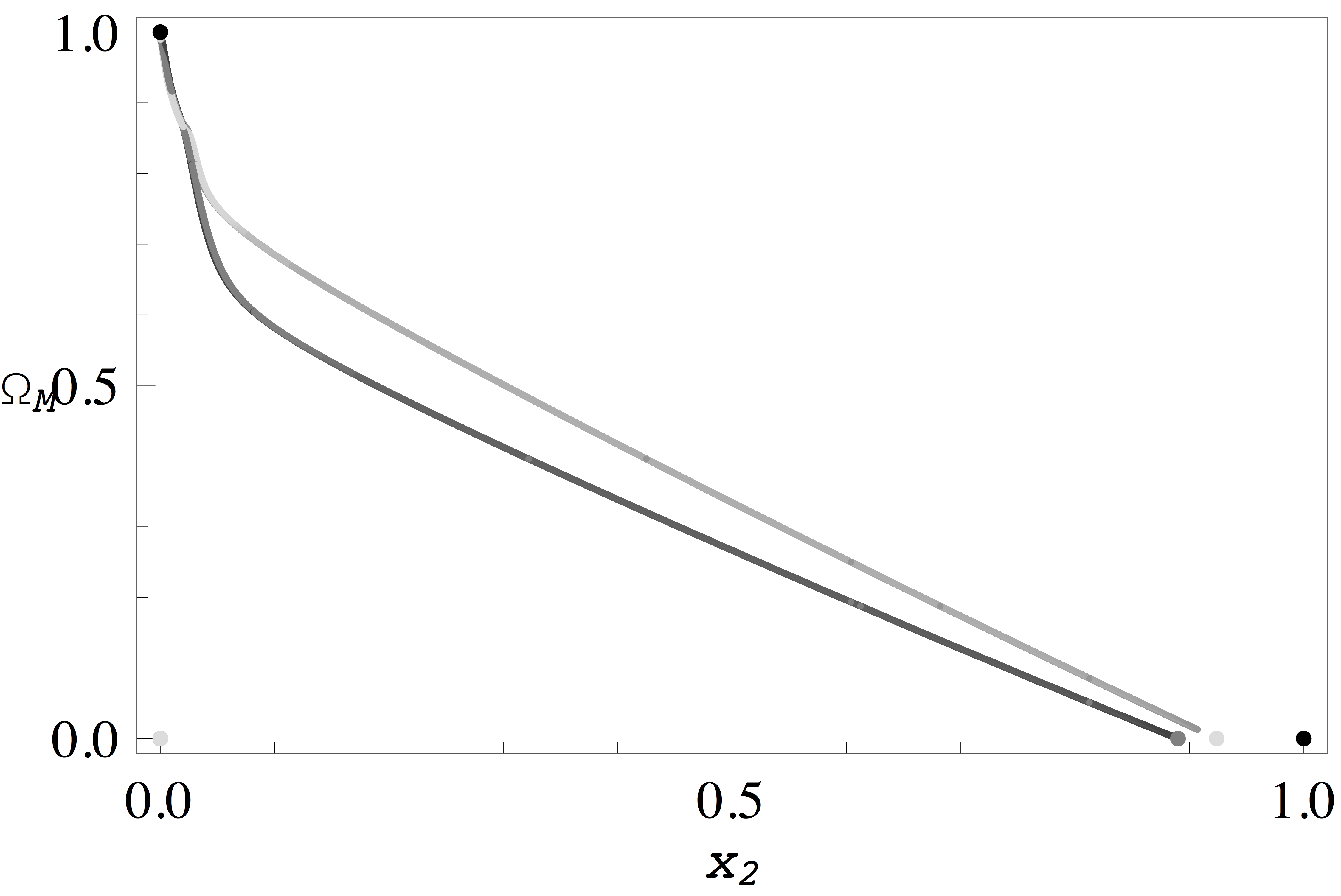}
\caption{Projection of the phase-space into the $x_2$-$\Om_M$ plane for $n=4$ (two darker plots) and $n=6$ (two lighter plots). Fixed points are marked as dots in dark grey ($n=4$) and light grey ($n=6$). For both cases $\xi b^2 = 10^{-2}$ and $x_{2i}=10^{-3}\;\text{(darker)},10^{-12}\;\text{(lighter)}$.}
\label{FIG2_24}
\end{figure}

Fig. \ref{FIG2_24} shows that the trajectories decrease monotonically with a physically meaningful, positive matter-density profile. Two linear phases are discernible: a very steep phase corresponding to the initial plunge of the deceleration parameter and consequent overshoot $q < 1$, and a less pronounced decrease during the compensation of that overshoot and evolution towards the final state $x_2=(1+\xi b^2)/( 1+2n)$.

\subsection{Example of system divergence}
The second case, for $n=2$, is depicted in Figs. \ref{FIG1_45}, \ref{FIG1_25} and \ref{FIG1_24}, and exhibits a divergence of the trajectories. In this case, the system's attractor is insufficient to counteract the repulsive effect of the repulsor ${\bf x}_A$ close to the system's initial values at the start of the simulation. 

\begin{figure}[!h]
\includegraphics[scale=0.4]{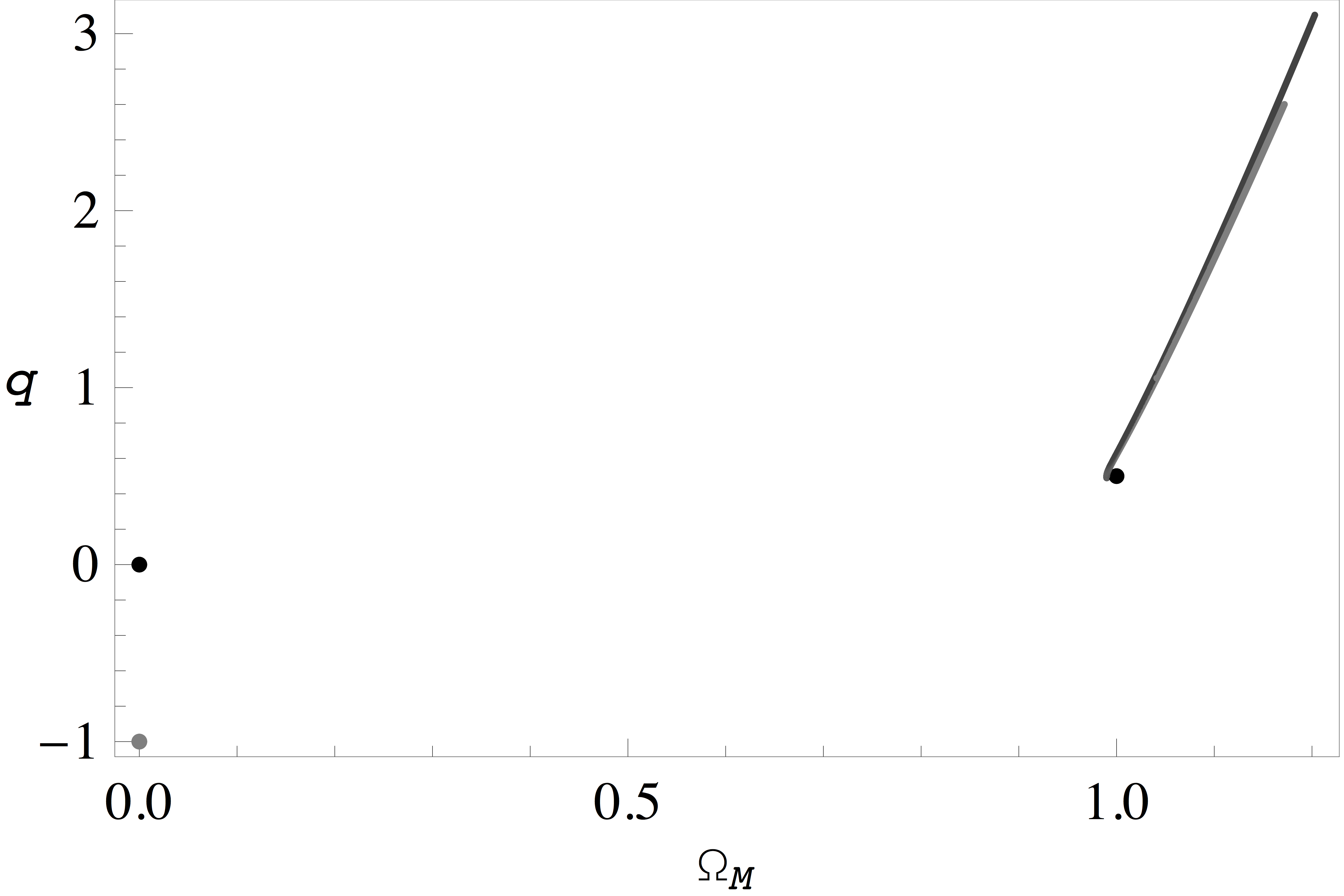}
\caption{Projection of the phase-space into the $\Om_M$-$q$ plane for $n=4$ (two darker plots) and $n=6$ (two lighter plots). Fixed points are marked as dots in dark grey ($n=4$) and light grey ($n=6$). For both cases $\xi b^2 = 10^{-2}$ and $x_{2i}=10^{-3}\;\text{(darker)},10^{-12}\;\text{(lighter)}$.}
\label{FIG1_45}
\end{figure}

\begin{figure}[!h]
\includegraphics[scale=0.4]{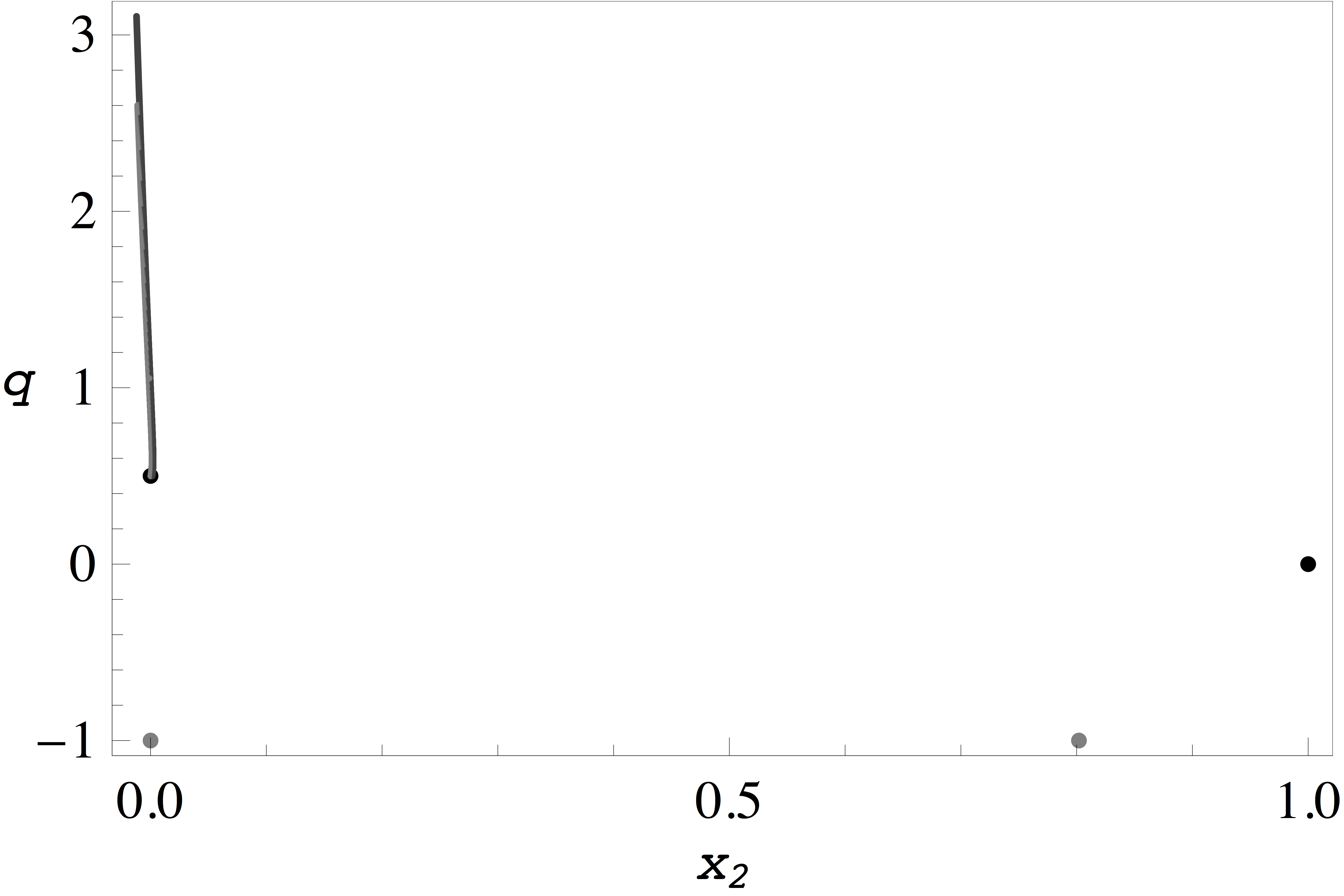}
\caption{Projection of the phase-space into the $x_2$-$q$ plane for $n=4$ (two darker plots) and $n=6$ (two lighter plots). Fixed points are marked as dots in dark grey ($n=4$) and light grey ($n=6$). For both cases $\xi b^2 = 10^{-2}$ and $x_{2i}=10^{-3}\;\text{(darker)},10^{-12}\;\text{(lighter)}$.}
\label{FIG1_25}
\end{figure}

In this case the analysis is much more straightforward: the fixed point ${\bf x}_A$ is so repulsive that it causes the trajectories to steer  clear of the close vicinity of the basin of convergence of the attractor, and therefore no physically significant convergence towards the latter is attained, as the trajectories lead the variables into physically meaningless values. Even if convergent trajectories for this case indeed exist, finding them constitutes largely a problem of fine-tuning.

\begin{figure}[!h]
\includegraphics[scale=0.4]{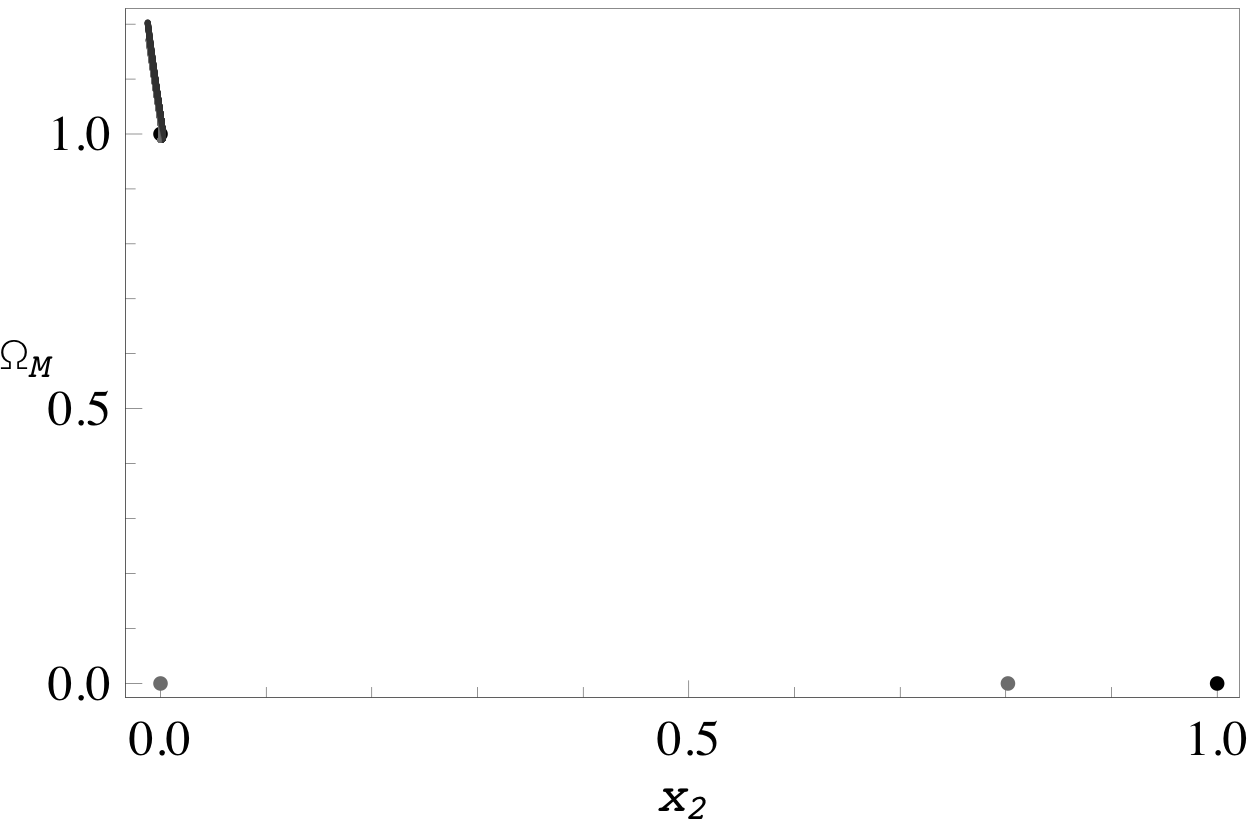}
\caption{Projection of the phase-space into the $x_2$-$\Om_M$ plane for $n=4$ (two darker plots) and $n=6$ (two lighter plots). Fixed points are marked as dots in dark grey ($n=4$) and light grey ($n=6$). For both cases $\xi b^2 = 10^{-2}$ and $x_{2i}=10^{-3}\;\text{(darker)},10^{-12}\;\text{(lighter)}$.}
\label{FIG1_24}
\end{figure}

\subsection{Overshoot variation}

As mentioned before, the analysed stable trajectories present a behavior for the deceleration parameter $q$ overshooting its final convergence value at the respective fixed point for the chosen potential. The magnitude of this effect, though it may appear to depend on the initial value taken for $x_2$, indeed flattens out after an initial phase (for approximately $-5.5 < x_{2i} < -2$) and responding linearly, eventually stabilising at a final value of $q_{Min} \approx -15.78$, for $n=4$, as can be seen in Fig. \ref{FIGovers_4}. 

\begin{figure}[!h]
\includegraphics[scale=0.4]{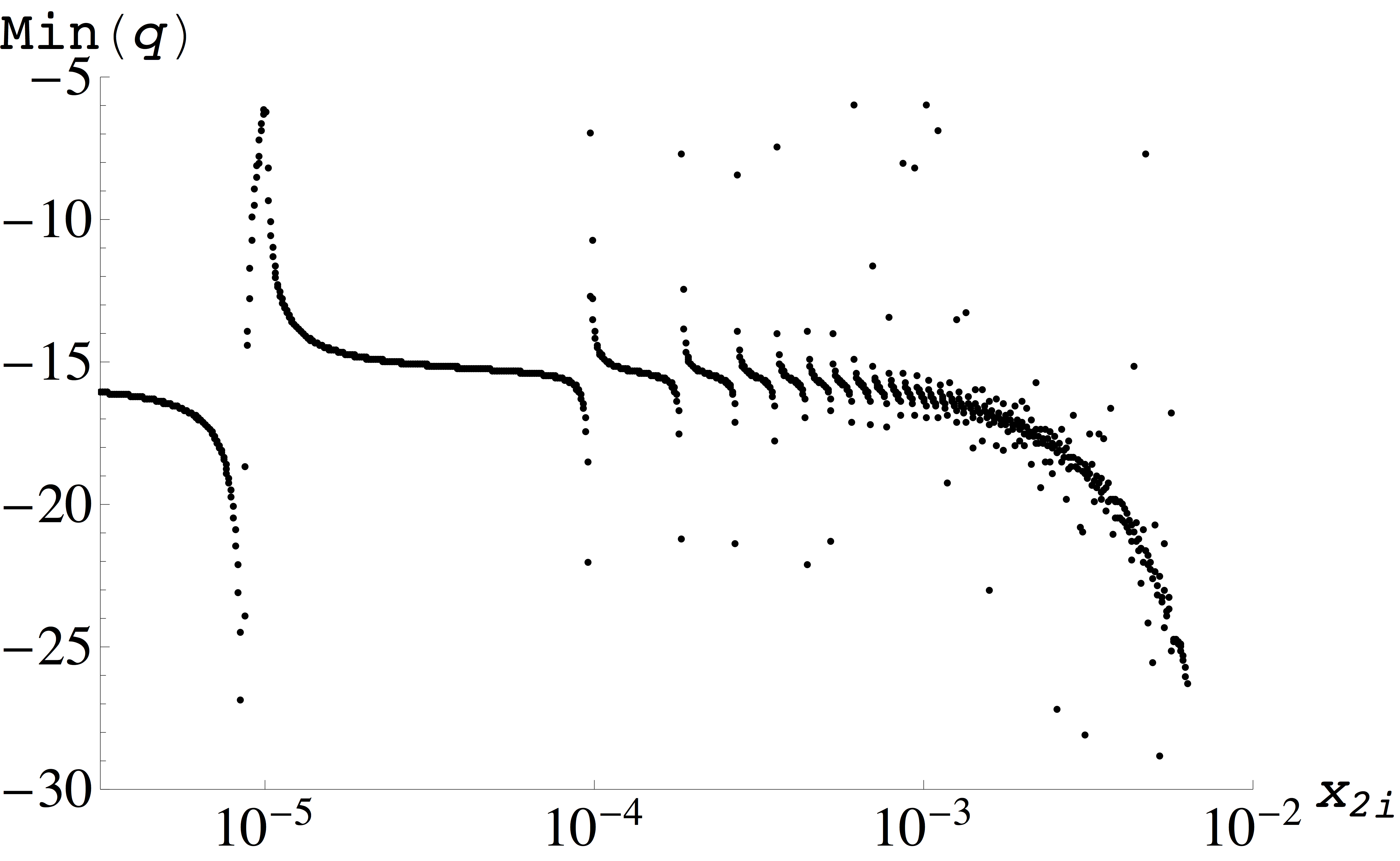}
\caption{Variation of the minimum of $q$ with the initial value of $x_2$ for $n=4$.}
\label{FIGovers_4}
\end{figure}

This result seems to suggest two zones of effective system evolution: the first one [$-5.5 < \log_{10}(x_{2i}) < \log_{10}(\xi b^2)$] more dynamical and vulnerable to the initial values, and the second one [$\log_{10}(x_{2i}) < -5.5$] more stable and robust to variations of the initial conditions.

As for the other studied case, $n=6$, presented in Fig. \ref{FIGovers_6}, it depicts a similar behavior, with the differences being a faster stabilization [a linear response is seen for $\log_{10}(x_{2i}) < -4.2$, leveling out at $q_{Min} \approx -20.47$], compared to when $n=4$.

\begin{figure}[!h]
\includegraphics[scale=0.4]{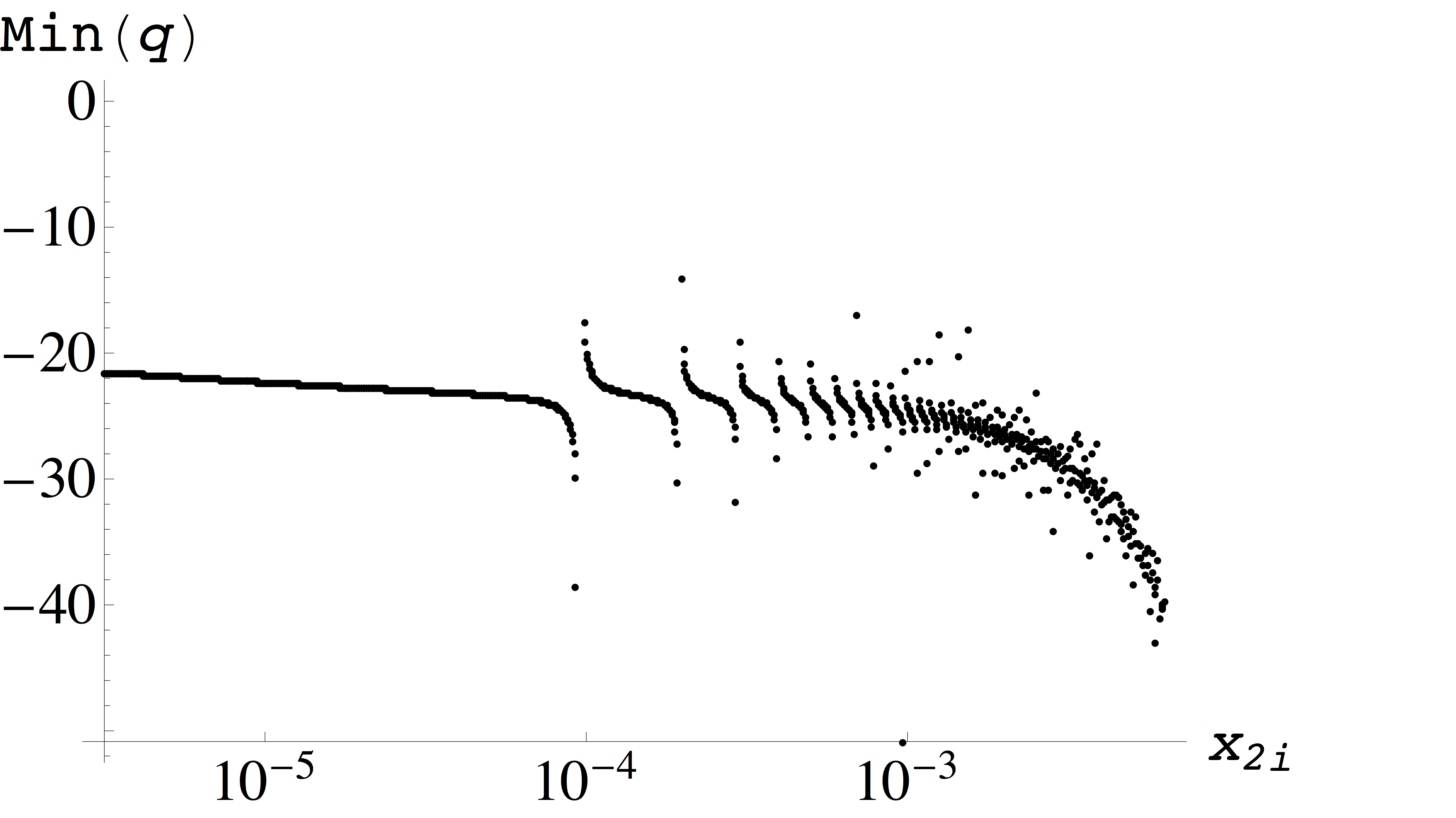}
\caption{Variation of the minimum of $q$ with the initial value of $x_2$ for $n=6$.}
\label{FIGovers_6}
\end{figure}

\section{Conclusions and Outlook}

In this work, one has considered the bumblebee model, a vectorial extension of general relativity under a spontaneous symmetry breaking mechanism. Since the study is made considering that the bumblebee vector is only nontrivial in the time component, which depends only on time, no additional breaking of Lorentz symmetry occurs, aside from the one derived from adopting the FRW metric, which is equivalent to a foliation of space-time into spatial pictures labelled by a cosmological time. 

It was found that the model predicts four points of equilibrium, in agreement with the preliminary study of the equations; two are unstable regardless of the value of the model's parameters, corresponding to the static and matter-dominated cases. The remaining two yield a Universe undergoing an accelerated expansion, which can be ascribed to the arise of a quantity akin to a cosmological constant: this can be caused by the effects of the potential alone (for practically all values of the parameters, with the exception of $-{1 \over 2}<n<0$), making the identification $\La \equiv 3 {H_0}^2$, or to the effects of a concerted evolution between the potential and the bumblebee vector's non-zero component, resulting in $\La^* \equiv -3{H_0}^2 \left( 1 + V_C \right) / 2$, where $V_C = (1\pm\xi b^2)/(1+2n)$ is the value of the potential on the fixed point ${\bf x}_C$. 

It was noted that, when considering a power-law type of potential, even within the parametrical constraints that guarantee the existence of one attractor, a convergent scenario isn't always found: in particular, this was not attained in the case of odd powers, which, for all tested values, appear to cause the system's variables to be strongly repelled initially, thus diverging from the de Sitter attractor. This can be explained by the fact that, for odd $n$, the variable $x_1$ will be initially negative if $x_2 < \xi b^2$ and positive if $x_2 > \xi b^2$. But the discussion concerning Eq. \ref{eqalpha} guarantees the system will diverge in the latter case, and in the former, the negativity of $x_1$ disturbs the balance of the Friedmann Equation (\ref{DINfried}), disallowing convergence.

The above does not guarantee that the system does not allow converging trajectories, only that they are not dense in the (continuous) set of all possible trajectories with physically significant starting points, and even if one was to be found, there would be no basis to argue the great fine-tuning needed for its manifestation.

Additionally, the analysis of convergent trajectories in phase-space predict that $q$ always overshoots the final value of $-1$, but since the present-day Universe is still on the descending slope of the trajectory, prior to that overshoot, such behavior constitutes a prediction of the model, rather than an experimental constraint (albeit one not to be verified in the foreseeable future).

Expectations concerning $\xi b^2$, $x_2$ and convergence of the system may, however, be used in restricting the window of accessible values to the dynamical system's variables: the major factors in that process can be categorized into three primary groups: (1) regularity of the solutions, (2) elimination of unphysical behavior and (3) mathematical stability of dynamical system.
 
\begin{enumerate}

\item $n<-1/2$ or $n>0$: This bound is required from a stability analysis of the fixed points of the dynamical system, as there are no converging trajectories for values of the exponent $n$ outside the indicated window. However, it can be argued that there is no guarantee that the evolution of the Universe must lead to stable convergence --- or that, even in the presence of suitable attractors, a region of converging trajectories can be found for a specific choice of model parameters $n$ and $\xi b^2$ without fine-tuning the initial conditions.

\item $x_{2i} < \xi b^2$:
After the spontaneous LSB has occurred and the potential has settled into its form (\ref{DINpot}), this upper bound for the value of the bumblebee field must be obeyed, otherwise the system diverges through the action of $\al$. 

\item $x_{2i}  \ll \xi b^2$: The stronger restriction, not contemplated in the Friedmann Equation, is related with the behavior of the deceleration parameter $q$, and ascertains that the minimum of the solution for this variable (the so-called overshoot) shows a tendency to explode if the absolute initial values of $x_2$ and $\xi b^2$ are too close. Once more, this is due to the dimensionless quantity $\al$ diverging and dominating the dynamic of the system, making impossible the smooth transition required in order for there not to be an overshoot.

\end{enumerate}

Regarding the obtained interval for the exponent $n$ where at least one of the de Sitter fixed points is an attractor (first point of the list above), the comparison with a more complex, nonpolynomial potential could provide interesting hints: indeed, if the latter can be expressed analytically as a series of powers of $B^2$, the two should match in the relevant regime for $B^2$ where the latter can be truncated.

Such a potential naturally arises from the asymptotically free solutions of a quantized bumblebee model \cite{SMEg3}, being expressed in terms of the Kummer (confluent hypergeometric) function $M(\ka-2,2,z) -1$. However, the stability analysis found there did not assess the dynamical (cosmological) evolution of the bumblebee field, but concentrated on the existence of minima of the potential, so that the former could rest on a nonvanishing VEV --- with excitations around the latter giving rise to Nambu-Goldstone modes associated with the ensuing LSB.

Our scenario differs significantly: it is classical in nature, has one nonvanishing minima defined by the power-law form of the potential, and the fixed points show that the bumblebee field does not evolve towards this nonvanishing VEV; furthermore, the results obtained in this study rely extensively on the presence of a coupling with geometry, which is absent from Ref. \cite{SMEg3}. As such, no direct comparison is feasible.

As a final remark, we note that the fairly rich structure of the model prompts more focused  studies: going beyond further inquiries into cosmological dynamics [{\it e.g.} by selecting a different potential $V(B^2 \pm b^2)$ or including a coupling between the bumblebee field and the Ricci scalar], one anticipates that interesting results should arise from a spatially oriented bumblebee that breaks the homogeneity and isotropy postulated by the cosmological principle: this could bear a possible relation with the putative ``axis of evil'' reported in the cosmological background radiation \cite{Bershadskii:2003rw}.

A different approach could also consider several local instances of spontaneous Lorentz symmetry breaking, in analogy with Condensed Matter Physics \cite{Kibble:2002ex}: one can argue that this is compatible with the above-mentioned possibility of a spontaneous LSB in a spatial direction, so that the bumblebee field acquires a particular (random) orientation only within a region with a well defined coherence length, beyond which another orientation is randomly adopted during the LSB. If the size of the observable Universe is much larger than the coherence length, it is physically plausible that the large scale superposition of several regions with different spatial orientations for the bumblebee field gives rise to a globally homogeneous and isotropic Universe. The collision of different ``bubbles'' could also produce interesting physics and hypothetically lead to observable relics of the LSB.

The in-depth study of such a mechanism would most probably require modeling the actual LSB ({\it i.e.} the evolution of the shape of the potential, so that the VEV of the bumblebee field becomes nonvanishing) and encompass similar considerations to the well-known Kibble-Zurek mechanism \cite{Kibble:2002ex}, albeit with the added complexity of considering a vector field instead of a scalar. By the same token, one remarks that the interface between regions with different spatial LSB could give rise to topological defects, which would physically correspond to areas where Lorentz symmetry holds.

However mesmerizing, the above scenario clearly requires heavy-duty numerical computations, and is thus incompatible with the stated purpose of this work: to clearly identify the possibility of driving an accelerated expansion of the Universe via a nonminimally coupled bumblebee vector field.

\acknowledgments

The authors thank O. Bertolami for fruitful discussion, and the referees for their useful remarks. D.C. was partially supported by the research grant Ref. BIC.UTL/Santander.05/12; J.P. was partially supported by Funda\c{c}\~ao para a Ci\^encia e a Tecnologia under the project PTDC/FIS/111362/2009.

\appendix

\section*{Additional bumblebee-curvature coupling}
\label{curv_coup}

As discussed in section \ref{cosmology}, the action functional (\ref{eqAction}) considered in this study follows Ref.  \cite{kostelecky1}, with only a coupling between the bumblebee field and the Ricci tensor considered: we argue that this captures the essential features of the model, namely the ensuing de Sitter attractor ${\bf x}_C$ depending on the coupling strength $\xi$.

If one additionally considers the coupling term in the action functional
\beq S = \int {\chi \over 2\ka} B_\mu B^\mu R \sqrt{-g} d^4 x~~, \eeq
\noindent where $\chi$ is a coupling strength (with dimensions $ [\chi] = [\xi ] =M^{-2}$), the modified bumblebee \eq{eqbumblebee} becomes
\beq
\nabla_\mu B^\mn = 2   \left( V' B^\nu - {\xi \over 2\ka}   B_\mu R^\mn - {\chi \over 2\ka} B^\nu R \right)~~,
\label{eqbumblebee_add}
\eeq
\noindent and the {\it r.h.s.} of the modified Einstein field equations acquires the extra terms
\beqa && - \chi [ B_\al B^\al G_\mn + R B_\mu B_\nu + \\ \nonumber && g_\mn \square(B_\al B^\al) - \nabla_\mu \nabla_\nu (B_\al B^\al) ]~~. \eeqa
Introducing the FRW metric (\ref{FRW}), \eq{eqbumblebee_add} becomes
\beq
\left[ V' - {3(2 \chi + \xi) \over 2\ka} {\ddot{a}\over a} - {3\chi \over \ka}H^2 \right] B = 0 ~~,
\label{eqbumblebee_add2}
\eeq
\noindent so that, considering the additional contribution to the derivative of the potential the curvature appearing on the {\it r.h.s.} of the  field equations, $ \De V' \equiv (\chi/2\ka)R$, the scalar curvature transforms as
\beq R \to (1-\xi B^2) R - 3\chi \square(B_\al B^\al) ~~, \eeq
The modified Friedmann and Raychaudhuri Eqs. (\ref{eqEinstTT},\ref{eqEinstRR}) thus read
\beqa \nonumber && H^2 [1- ( \xi + \chi ) B^2] = {1 \over 3} \ka (\rho + V ) + ( \xi + 2 \chi ) H B \dot B ~~,\\ && \nonumber \left( H^2  + 2 {\ddot{a} \over a} \right) [1- ( \chi + \xi ) B^2] = \\ && \ka V + 4( \xi + \chi ) HB\dot B + ( \xi + 2\chi ) ( \dot B ^2 + B \ddot B )~~.
\label{eqEinst_add} \eeqa
One cannot absorb the effect of the additional coupling into a redefined parameter, as both factors $\xi + \chi$ and $\xi + 2\chi$ appear; notwithstanding, the above shows that the dynamics do not differ significantly from those obtained with $\chi = 0$ --- as no additional combinations of $H$, $B$ or its derivatives arise.

One can apply the procedure followed in Section \ref{SSECdesitter} to ascertain the possibility of obtaining a de Sitter phase from this more evolved scenario where two couplings between the bumblebee field and geometry. The modified bumblebee \eq{eqbumblebee_add2} becomes
\beq V' = {3(4 \chi + \xi) \over 2\ka} H_0^2 ~~, \eeq
\noindent again providing an intrinsic relation setting $B(t) = B_0 = {\it const.}$. Introducing this relation into \eq{eqEinst_add} yields
\beq \label{eqHVBsitter_add} H_0^2 = {\ka V_0 \over 3[1-(\xi + \chi) B_0^2]} ~~~~,~~~~ \rho = 0 ~~, \eeq
\noindent which, as before, describes a Universe with no matter content and with the potential term acting as a dark energy component driving its accelerated expansion. Interestingly, the above shows that in the particular case $\xi = - \chi$ the two couplings cancel.

\end{document}